# The Ancient Astronomy of Easter Island: Kirch's Comet

Sergei Rjabchikov[1]

[1]The Sergei Rjabchikov Foundation - Research Centre for Studies of Ancient Civilisations and Cultures, Krasnodar, Russia, e-mail: srjabchikov@hotmail.com

**Abstract**

A fragment of the folklore text "*Apai*" that was once put down on Easter Island contains a report about astronomical observations of Kirch's Comet or the Great Comet of 1680-1681 A.D. (C/1680 V1) and the partial solar eclipse of March 19, 1681 A.D. Some astronomical aspects of the local cult of bird-men have been discussed, too.

**Keywords**: archaeoastronomy, writing, folklore, rock art, Rapanui, Rapa Nui, Easter Island, Polynesia

## Introduction

The civilisation of Easter Island is famous due to their numerous ceremonial platforms oriented on the sun (Mulloy 1961, 1973, 1975; Liller 1991). One can therefore presume that some folklore sources as well as *rongorongo* inscriptions retained documents of ancient priest-astronomers.

## A Brief Report in a Rapanui Folklore Text

In the folklore text "*Apai*" (an oral version of a lesson book in the royal *rongorongo* school) taken down on Easter Island (Thomson 1891: 517-518) there is a passage:

*Ui te taura hiku raverave. A Hiro kai te teri he po.* 'A priest *taura* stared at a long tail (*hiku raverave*). (Afterwards the chthonic god) *Hiro* ate up the sun (*teri*) before a certain night.' (It is my own translation.)

## The Linguistic Background

In this quotation the term *teri* is obscure only; on the base of the analysis of the text it is the designation of the sun. I have demonstrated that Rapanui *renga* 'yellow; beautiful' was its epithet (Rjabchikov 1998); as a result the terms *teri* or *tero* (the sun) because of the gradations of the sounds *i* and *o* are comparable with Mangarevan *tero* 'handsome.' Let us examine the inscriptions on two tablets which contain this archaic word.

1. Consider a record on the Tahua (A) tablet that served as a *rongorongo* lesson book, see figure 1.

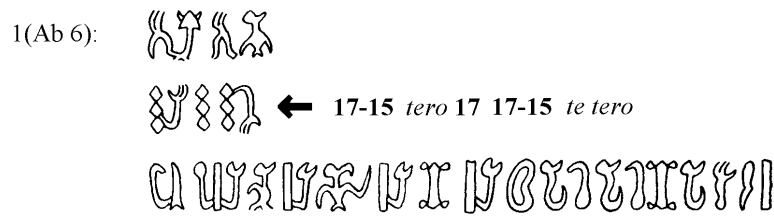

Figure 1.

1 (Ab 6): **2** (= a reversed variant) **21 2** (= a reversed variant) **44** *Ina oko, ina taha.* 'There is no ripeness; there is no food.'
**17-15 17 17-15** *Tero, te tero.* [or: *Teri, te teri.*] 'The brilliance of the sun increases.'
**35 5** *Patu:* '(Yam) leaves appear:'
**4-4 27 19** *atuatu rau Kio(r)e,* 'the variety-leaves (called) *Kioe*,'

**4 27 44** *atu rau Taha,* 'the variety-leaves (called) *Taha,*'
**4 27 56-56** *atu rau Papa,* 'the variety-leaves (called) *Papa,*'
**4 27 47** *atu rau (R)ava,* 'the variety-leaves (called) *Ravei,*'
**27 25** *rau (R)ua,* '(the variety)-leaves (called) *Rua,*'
**27 25 56-56** *rau (R)ua papa,* '(the variety)-leaves (called) *Rua papa,*'
**27 15** *rau roa.* '(and the variety) with large leaves [*Apuku raurau*].'
**50-4** *Hitu.* '(They all) are seven (varieties).'

    This record has been partially decoded (Rjabchikov 1993a: 138-139, appendix 2, figure 5, fragment 66). It tells of the planting and growth of yams in the spring-time. Old Rapanui *taha* 'portion of food; food' is cognate with Mangarevan *tahatu* 'portion of food yet remained,' Maori *taha* 'portion' and Samoan *tafa* 'to cut.' Rapanui yam varieties have been well documented (Métraux 1940: 155; Barthel 1978: 108-110). Cf. also Hawaiian *pālau* [< *pa rau] 'yam; a variety of yam.'

    2. Consider three parallel records on the Aruku-Kurenga (B) tablet that served also as a *rongorongo* lesson book, see figures 2 to 4. These records have been partially decoded (Rjabchikov 1995). Fedorova (1982) at first assumed that glyphs **7-70** denoted the star Aldebaran (α Tauri). Nonetheless, later on she (Fedorova 2001) gave this truly interpretation up. Besides, Lee (1992: 80) believes that in some instances the turtle can be displayed as an emblem of the Pleiades (M45; NGC 1432) in the Rapanui rock art. Although I offered at the outset that the turtle glyph **68** *honu, hono, ono* denoted the whole constellation of Taurus in some contexts in *rongorongo* inscriptions (Rjabchikov 1993b: 5, table 1), now I am inclined to suggest that this glyph could, on occasion, represent the Pleiades only.

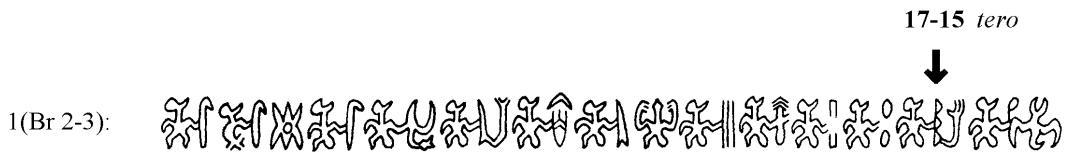

Figure 2.

**6-35-6-35 7 6-35 2 … 6 9 33 … 6 17-15 6 40var 81**
*(H)apa(h)apa tuu, (h)apa Hina ... a niva ua … a Tero* [or: *Teri*], *a Re Manu.*
'The star rose, the Moon goddess, … the darkness because of rains … the sun, the Flying (*Rere*) Bird (appeared).'

    Here the star (*tuu*) is Aldebaran (see below). Its emergence before dawn in the rainy season is described. The natives waited for the time of the increasing hot of the sun as well as the arrival of the sooty terns in September.

    Let us study a folklore text about the god *Makemake* (Felbermayer 1971: 41-44). It contains the following verb *apa* 'to lift' that was not documented in the Rapanui vocabularies: *He apa no te toe i te vave nunui i te tomohanga.* 'The big waves (*te vave nunui*) rose (*apa*) higher (*no te toe*) during (his) landing.' (It is my own translation.) Instead of the terms *apai* and *hapai* 'to lift' (< *sapai < *sapa-'i*), their more archaic version, *(h)apa* 'ditto' (< *sapa*), is presented. Old Rapanui *niva* 'darkness' fits Maori *niwaniwa* 'dark, deep black.'

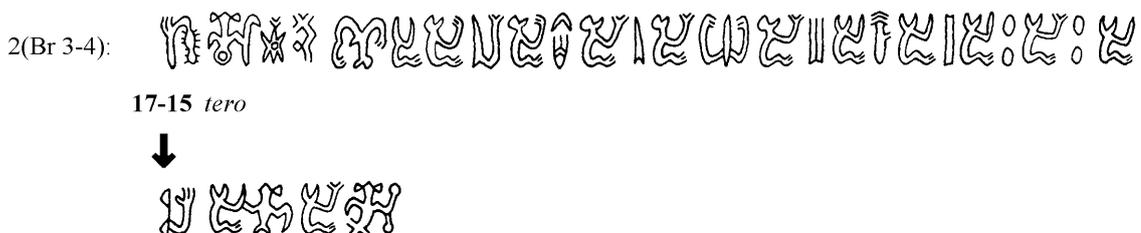

Figure 3.

**15-25 68/65 35 7-70 31 2 … 6 9 33 … 6 17-15 6 81 6 6-4**

*Rohu Honu RANGI, pe Tuu Pu Make(make); Hina … a niva ua … a Tero* [or: *Teri*], *a Manu, a Hatu.*
'The Pleiades IN THE SKY, (and then on the same path) Aldebaran were created before dawn (= the sun god *Make-make* literally); the Moon goddess, … the darkness because of rains … the sun, the Bird, (the sun god) *Tiki-te-Hatu* (appeared).'

In two Rapanui rock drawings (Lee 1992: 102, figure 4.97; 151, figure 5.24) at the ceremonial centre Mata Ngarau, Orongo, I have disclosed glyph **7** *tuu* denoting the star. Such symbols rendered in the midst of the petroglyphs of sacred birds or bird-men identify in all likelihood Aldebaran.

**17-15** *tero*
↓

3(Br 6): 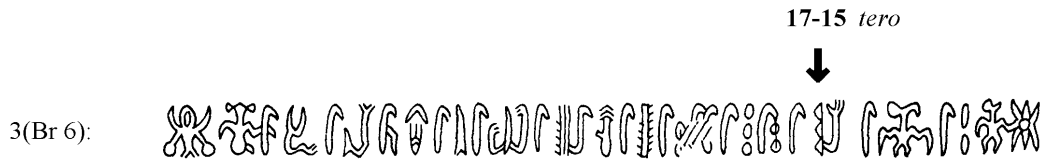

Figure 4.

**79 68 35 2 … 35 9 33 … 35 17-15 35 44 35 29 6-7**
*Eke Honu; pe Hina … pe niva ua … pe Tero* [or: *Teri*], *pe Ta(h)a, pe Rua Hatu.*
'The Pleiades rose; toward the Moon goddess … toward the darkness because of rains … toward the sun, toward the Frigate Bird (the solar symbolism; in some cases this glyph reads *manu tara* figuratively and represents the sooty tern), toward (the [bird] character) *Rurua-Tiki-te-Hatu* (appeared).'

In one Rapanui rock drawing (Lee 1992: 150, figure 5.22) at the ceremonial centre Mata Ngarau, Orongo, I have disclosed among the symbols of bird-men glyph **68** *honu*, *hono*, *ono* (cf. Rapanui *ono* 'six') denoting here the Pleiades. The natives saw in the sky this small cluster counting six or seven stars. Métraux (1940: 323) has correlated the name of the female personage *Rurua-Tiki-te-Hatu* with the Tuamotuan mythological name *Rua-Tiki*.

On the strength of PMP *\*dilap* and *\*dilep* 'to shine' (ACD) one can reconstruct the evolution of a derivative of these forms: *\*dil- > \*del- > \*tel- > \*ter-*.

**On the New Zealand Designations of Some Comets as Celestial Tails**

Best (1922: 53, 56) says that *Taketake-hikuroa* was the Maori name of a comet. The last components of it are *hiku roa* 'long tail.'

Another Maori term associated with comets was *hiku makohurangi* 'misty tail' (Ibid., p. 55); this expression [*hiku ma-kohu rangi*] literally signifies 'misty tail in the sky.'

A further Maori term treated on a comet was the name of the mythological Mount Hikurangi (Ibid., p. 13), literally meaning [*Hiku rangi*] 'Tail in the sky.'

**The Ultimate Interpretation of the Passage of the "*Apai*" Text**

For years 1600 – 1860 A.D. I have calculated all the dates of the solar eclipses on Easter Island using the computer program RedShift Multimedia Astronomy (Maris Multimedia, San Rafael, USA). I have taken into account the information about comets that were visible by the naked eye during that time period (Kronk 1999; 2003).

As one would expect, one solution of our problem has been obtained. The long tail (*hiku raverave*) was Kirch's Comet or the Great Comet of 1680-1681 A.D. (C/1680 V1) in reality. The narration about the eaten sun (*kai te teri*) during its setting (*he po*) was a mythological sketch (cf.: *a Hiro*) of the partial solar eclipse of March 19, 1681 A.D.

**Conclusions**

So, the ancient Rapanui report concerning a large comet and a solar eclipse has been interpreted successfully. The deciphered fragments on the Aruku-Kurenga tablet as well as the corresponding local rock drawings witness that the local priest-astronomers watched Aldebaran and the Pleiades to determine their positions as celestial markers of the appearance of the sooty terns in September.